\documentclass[aps,pra,groupedaddress,amsmath,amssymb,amsbsy,twocolumn]{revtex4-1}
\usepackage{graphicx}
\usepackage{caption}
\usepackage{amsmath}
\usepackage{color}
\usepackage{epstopdf}
\begin{document}

\title{Full statistical mode reconstruction of a light field via a photon-number resolved measurement}

\author{I.A. Burenkov$^{1,2}$, A.K. Sharma$^2$, T. Gerrits$^3$, G. Harder$^4$, T.J. Bartley$^{3,4}$,  C. Silberhorn$^4$, E.A. Goldschmidt$^5$ and S.V. Polyakov$^2$ }
\address{$^1$ Joint Quantum Institute \& University of Maryland, College Park, MD 20742, USA.\\
         $^2$ National Institute of Standards and Technology, Gaithersburg, MD 20899, USA.\\
         $^3$ National Institute of Standards and Technology, Boulder, CO 80305, USA\\
         $^4$ University of Paderborn, 33098 Paderborn, Germany\\
				 $^5$ United States Army Research Laboratory, Adelphi, MD 20783, USA}
\email{ivan.burenkov@gmail.com}

\begin{abstract}
We present a method to reconstruct the complete statistical mode structure and optical losses of multimode conjugated optical fields using an experimentally measured joint photon-number probability distribution. We demonstrate that this method evaluates classical and non-classical properties using a single measurement technique and is well-suited for quantum mesoscopic state characterization. We obtain a nearly-perfect reconstruction of a field comprised of up to 10 modes based on a minimal set of assumptions. To show the utility of this method, we use it to reconstruct the mode structure of an unknown bright parametric down-conversion source. 
\end{abstract}
\maketitle
%\ocis{270.5290, 270.6570.}
\section*{Introduction}
Macroscopic  and mesoscopic quantum phenomena, processes that exhibit quantum behavior at the large scale, have attracted the interest of researchers for a long time both for fundamental and practical applications \cite{MesoscopicBook, Hall, BEC}. The physics of such systems is at the forefront of modern physical science and is a subject of interest and research to a broad range of fields and related disciplines. A recurring theme behind mesoscopic physics phenomena is that methods successfully used in the microscopic and macroscopic world fail for this intermediate regime, calling for a whole new suite of approaches \cite{MesoscopicBook}. Mesoscopic {\it quantum} systems stand alone, because their behavior cannot be considered classical and their nonclassical properties are important but typically subtle, and hard to characterize \cite{Schmied441, Bianconi}. Yet, implementation of large quantum systems remains a difficult task, from the synthesis and manipulation of quantum states of an appropriate size to the measurement of their components. 
Much of this work is done in the optical domain, because optical states naturally offer low decoherence, and because nonlinear processes, such as parametric down-conversion (PDC) \cite{Brida,Silberhorn,PDC1,PDC2,PDC3} and four-wave mixing (FWM) \cite{FWM}, may produce conjugated fields (CFs) \cite{TB1,TB2, GT}, giving exactly the same number of photons in two arms, that are sufficiently scalable. Complete statistical mode structure characterization would reveal physical properties of such quantum sources, aiding their development. Unfortunately, traditional characterization measurements, which are well-suited for low photon-number states \cite{Clauser1974,Kwiat1991,Chou2004} are of little use for brighter states, as shown in recent publications \cite{Waks2006,Christ2011,Dyakonov2015}. Therefore, new tools for studying mesoscopic systems are of broad interest. Beyond the immediate appeal to quantum optics and scalable quantum information processing, such methods can be applied to complex systems analysis \cite{Bianconi}, machine learning \cite{Saul1999, Amati}, Bose-Einstein condensates \cite{Schmied441} and other systems in different fields of science.
%One consideration, increasingly important for bright states of light, is their complete mode structure, which fully determines their properties: from their interaction with other quantum states of light to their detection. In addition, the mode structure contains rich information about the source of a light field, and its scalability potential. 

In the current work, we solved a long-standing problem of linking the directly observable statistical properties of quantum states to the physical properties of these states. To our knowledge, our method is the first characterization method that %scales favorably with the brightness of the state under measurement. , while the method introduced in the previous cannot be extended to multiple beams, as in that case its reconstructions become unstable and mesoscopic states cannot be analyzed. In contrast, our current approach 
scales favorably with the state brightness, works well with mesoscopic states, and can be used to assess their nonclassicality \cite{Hillery}. %We point out that to our knowledge there exists no other characterization method that assesses the quantum properties of bright light.
Because optical modes contribute to the photon number statistics, the mode structure of conjugated sources of light can be reconstructed using a photon-number resolved joint probability distribution (JPD). It identifies their basic components, a set of correlated and uncorrelated optical modes, thus enabling {\it in situ} characterization and remote sensing. 
We numerically demonstrate the uncertainty reduction in the mode reconstruction algorithm with the source brightness. Additionally, this method identifies the overall optical losses for conjugated fields, an otherwise difficult task in many cases where there is loss associated with the pair production medium itself \cite{Vahlbruch2016}. This method is loss-tolerant and thus is directly applicable to realistic mesoscopic and macroscopic quantum states of light. 
Most notably, for the first time, we demonstrate how to determine the mode structure of quantum mesoscopic states in a single measurement and with a minimal set of assumptions, i.e. how to identify the number and statistical types of the modes from the same dataset. 
Photon number resolved detection up to nearly 100 photons per pulse has been recently demonstrated \cite{GT,PNR}. It is an excellent tool for a JPD measurement. 
Using our method, we identified the mode structure of a bright PDC source, and demonstrated its nonclassicalty, using only an experimentally obtained JPD.

%Most notably, mode structure can be acquired without direct access to the source, i.e. merely by studying Glauber coherence functions of remotely detected light \cite{PolyGold}

% We demonstrate how to identify the number and types of modes.  

%Using the measured statistics of a mesoscopic system we are able to extract its physical properties that are difficult to determine using the methods typically applied to micro and macroscopic systems. In particular, this method recovers quantum properties that are not accessible by any other method for this range of scales. The growing interest in scalable quantum computers makes the development of tools for studying of mesoscopic quantum states all the more pressing. %In addition, this is the first scalable method capable of assessing mesoscopic quantum states of light.

\section*{Mode reconstruction method}

Consider a JPD for two CFs in the general case. The two arms are denoted `signal' (s) and `idler' (i), respectively. They are comprised of one or more optical modes. There are perfectly conjugated modes that are generated as photons in pairs with a particular statistical distribution and independent losses for each mode in each arm. There are also uncorrelated fields in each arm. We write the total JPD $P(n_\mathrm{s},n_\mathrm{i})$ in terms of underlying distributions. Here $n_\mathrm{s}$, $n_\mathrm{i}$ are the number of photons detected in the signal, idler arms respectively, and the underlying modes have probability distributions $p_{\mu}(n)$ for mean photon numbers $\mu$. $L_{n,k}(\eta)=\eta^n(1-\eta)^{k-n}k!/((k-n)!n!)$ are loss probability factors (LPF)s that compute a probability that $n\leq k$ photons are measured given transmittance $\eta$ and $k$ initial photons. This loss model enables reconstruction of light sources comprised of modes with any statistics (see Appendix A). 
For all optical mesoscopic sources demonstrated to date, losses in an uncorrelated mode will only affect the mean photon number in that mode while leaving the statistics unaltered. Using this fact, we write the uncorrelated part of the JPD with loss-adjusted $\tilde{\mu}_j=\mu_j\eta_j$, thereby significantly simplifying the evaluation of probabilities, see Appendix A for details. 
Then the JPD $P(n_\mathrm{s},n_\mathrm{i})$ can be calculated as:
%\begin{widetext}
\begin{equation}
\begin{split}
P_\mathrm{c}(N_\mathrm{s},N_\mathrm{i})=&\overset{\infty}{\underset{(N_\mathrm{s}, N_\mathrm{i})}{\underset{\mathrm{Max}}{\underset{k=}{\sum}}}}\underset{\Sigma n^\mathrm{i}_j=N_\mathrm{i}}{\underset{\Sigma n^\mathrm{s}_j=N_\mathrm{s}}{\underset{\Sigma k_j=k} {\sum}}} \underset{j}{\prod}p_{\mu_j}(k_j)L_{n^\mathrm{s}_j,k_j}(\eta^\mathrm{s}_j)L_{n^\mathrm{i}_j,k_j}(\eta^\mathrm{i}_j)  \\
P_\mathrm{u}(M)=&{\underset{\sum{k_j}=M} {\sum}} \underset{j}{\prod}p_{\tilde{\mu}_j}(k_j)\\
P(n_\mathrm{s},n_\mathrm{i})=&\underset{N_\mathrm{i}+M_\mathrm{i}=n_\mathrm{i}}{\underset{N_\mathrm{s}+M_\mathrm{s}=n_\mathrm{s}}{\sum}} P_\mathrm{c}(N_\mathrm{s}, N_\mathrm{i})P_{u_\mathrm{s}}(M_\mathrm{s})P_{u_\mathrm{i}}(M_\mathrm{i})
\label{eqn:JPD}
\end{split}
\end{equation}
%\end{widetext}
where $P_\mathrm{c}$ and $P_\mathrm{u}$ are correlated and uncorrelated parts of the JPD.

We consider three photon number probability distributions that cover a broad range of conjugated sources, including those based on PDC and FWM. These are thermal modes governed by Bose-Einstein statistics: $p_{\mu}^{\rm Therm}(k)=\mu^k/(1+\mu)^{k+1}$, Poissonian modes: $p_{\mu}^{\rm Pois}(k)={\rm exp}(-\mu)\mu^k/k!$, and single-photon modes governed by binomial statistics: $p_{\mu}^{\rm SP}(0)=(1-\mu)$; $p_{\mu}^{\rm SP}(1)=\mu$; and $p_{\mu}^{\rm SP}(k>1)=0$. Therefore, a pair of CFs can be described by the parameter set: $\mathcal{S}=\{\{$type$_j$, $\mu_j$, $\eta^\mathrm{s}_j$, $\eta^\mathrm{i}_j\}_c;$ $\{$type$_j$, $\mu_j\eta^\mathrm{s}_j\}_s;$ $\{$type$_j$, $\mu_j\eta^\mathrm{i}_j\}_i\}$ for all the modes that may be present in the system. Here, type indicates a mode (`Therm', `Pois', and `SP' for thermal, Poissonian and single photon modes respectively) and subscripts refer to field occupancy (`s' for signal only, `i' for idler only and `c' for conjugated). For each mode, losses do not change the statistics, only the mean photon number $\mu$. Note that in general, different modes may experience different losses due, for instance, to imperfect spatial overlap between emitted and collected light. We are interested in finding the full set of mode parameters based on an experimental measurement. Ordinarily, a nonlinear parametric fit could be employed on a measured JPD, to minimize $\sum\left[(\sqrt{x_j}-\sqrt{f_j})/\sigma_j\right]^2$, a typical scoring function, where $x_j$ is the original vector, $f_j$ the fit vector, and $\sigma_j$ the uncertainties vector. In general, a large number of free parameters results in ambiguous fits.

A very important observation is that each of the two CFs can be assessed separately. Remarkably, no additional measurements are required for this analysis. One dimensional reduced probability distributions (RPDs) for each field can be obtained from a JPD by a simple summation over rows and columns. Clearly, 1D RPDs fully describe modes that are present in each of the two conjugated fields. In addition, because the accuracy of a measured probability distribution is dominated by statistical uncertainty, summing over rows or columns of the 2D JPD means the 1D RPD has much lower statistical uncertainty. 

%An RPD can be directly used Mode reconstruction of a single light field is discussed in detail in our previous work \cite{PolyGold}. Based on this work, we build an appropriate parameter set with no prior knowledge of the mode types, see Supplementary information. 
Reconstruction of RPD is based on photon number distribution given by $P_\mathrm{u}(M)$ from (\ref{eqn:JPD}). The modes that are identified through an RPD analysis unambiguously define a mode structure of the overall CFs to within transmittance losses. In addition, modes cannot be designated as correlated or uncorrelated through an RPD reconstruction. Thus, an RPD reconstruction provides the two parameter sets $\{$type$_j$, $\mu_j\eta^\mathrm{s}_j\}$, $\{$type$_j$, $\mu_j\eta^\mathrm{i}_j\}$. 
These parameter sets significantly simplify the reconstruction of a JPD. First, the number and types of modes are fixed. Second, the overall number of fitting parameters for a 2D fit is reduced. Thus, we establish a two-step reconstruction method particularly suitable for the experimental use, i.e. when the number and types of modes are unknown.
Note that alternative approach based on Glauber coherence functions was considered in \cite{PolyGold} for the mode reconstruction of a single light field. However, while the two approaches are equivalent for dim states, they differ significantly for mesoscopic and bright states. For bright states, Glauber functions of non-classical states become practically indistinguishable from classical states with similar brightness. Thus, in our work, we use the advantage of statistical mode analysis based on a probability distribution.  

\section*{Reconstruction of simulated JPD$\mathrm{s}$}

To test this method, we calculate JPDs for typical multimode CF sources with known parameters, $\mathcal{S}_\mathrm{simulated}$. Then we find $\mathcal{S}_\mathrm{recovered}$ using our two-step reconstruction method and compare it to the initial parameter set, $\mathcal{S}_\mathrm{simulated}$. 

\begin{figure}
\centering
   \includegraphics[width=0.46\textwidth]{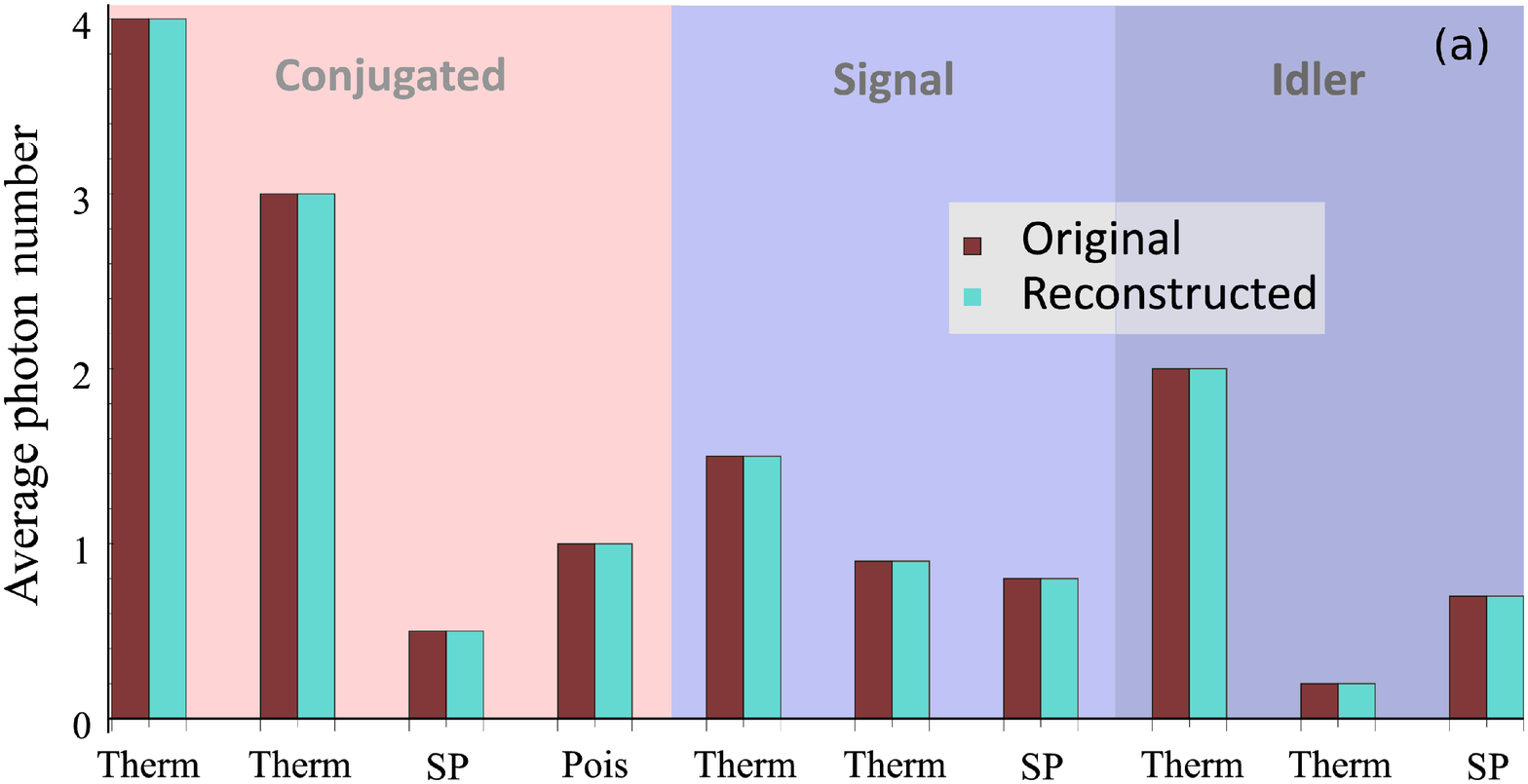}\hspace{0.05\textwidth}
   \vspace*{1cm}
   \includegraphics[width=0.45\textwidth]{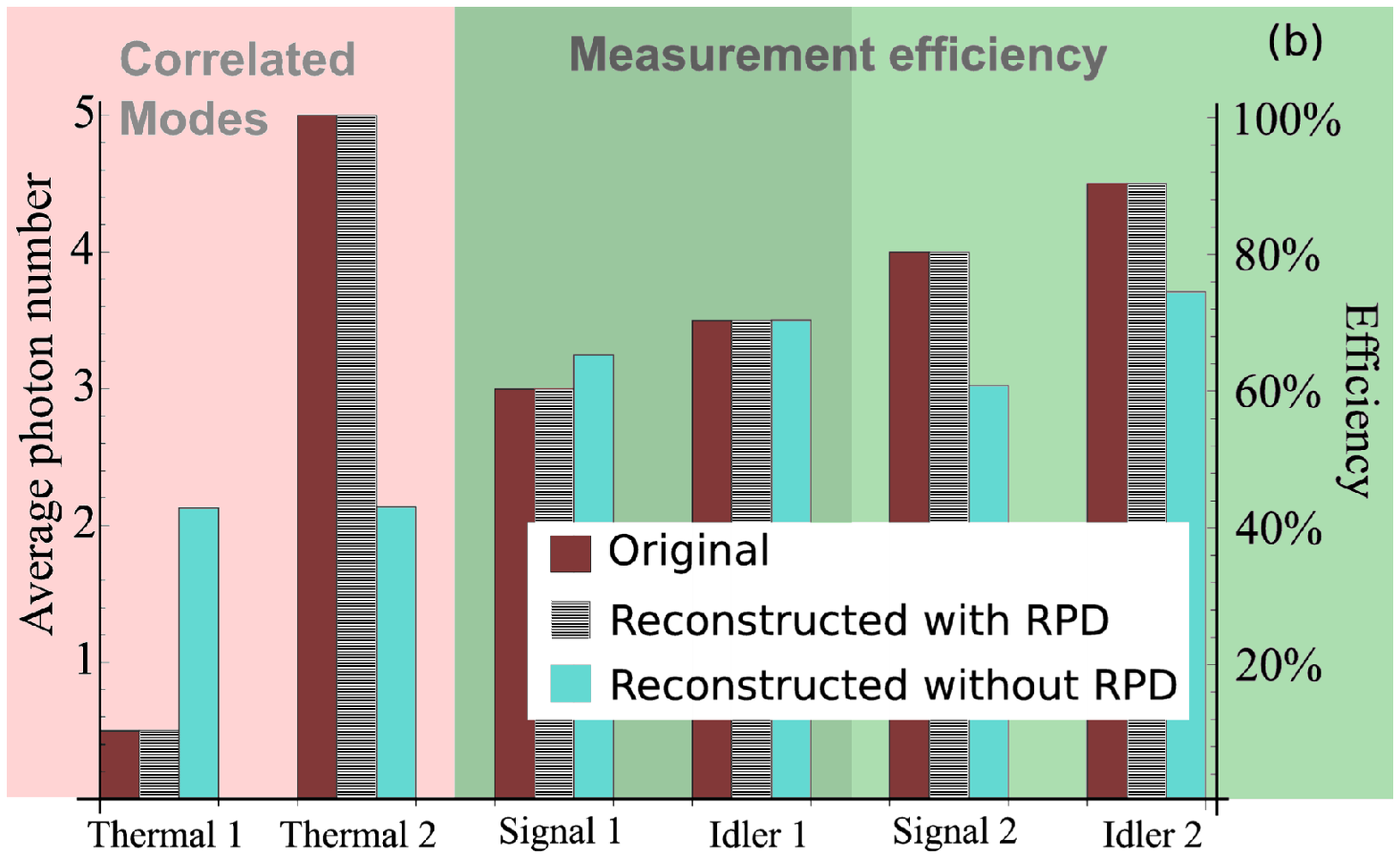}
\caption{\label{fig:one} A comparison of a simulated original and a reconstructed mode distributions for a typical conjugated source. (a) A source with 4 correlated and 6 uncorrelated thermal (Th), Poisson (P) and single photon (SP) modes (3 per channel signal/idler), showing an exact reconstruction. (b) A source with 2 correlated thermal modes and unknown independent losses reconstructed using unconstrained and constrained JPD fits. An unconstrained fit converges to a local minimum, while a two-step fit presents an exact reconstruction (see text).}
\end{figure}

We obtain perfect fits for fields with up to 10 modes using a simulated JPD both with no losses and with realistic transmittance losses. Such a perfect reconstruction example is shown in Fig. \ref{fig:one}(a). This is satisfying, but not surprising, because the probabilities in the JPD are computed exactly, using Eqs. (\ref{eqn:JPD}). For the general case of unknown losses, which can be different for different modes, the use of a two-step method with RPD reconstructions followed by a JPD reconstruction is essential, because scoring functions based on 2D JPDs in the most general case may suffer from multiple local minima. The advantage of defining number and types of modes with RPD fits is illustrated in Fig. \ref{fig:one}(b). In this simulation, two conjugated modes with two independent losses (each mode suffers different losses) in both signal and idler arms cannot be reconstructed without an initial RPD step. With this step we are able to reproduce not only the mean photon numbers in the correlated modes, but also the four independent loss coefficients. This is an important feature for using this method to optimize mode-matched collection from a multimode photon pair source \cite{Dixon2014}.

This method is limited mainly by the maximum photon number detected and the total amount of data accumulated. The total amount of data accumulated sets a shot noise level for that data, which directly affects the accuracy of any reconstruction. The highest resolved  photon-number state limits the number of modes that can be reconstructed. This is particularly important for determining the presence of a Poissonian mode rather than several similar thermal modes. Naturally, the source brightness {\it improves} both the reconstruction accuracy and the maximum possible number of reconstructed modes. In the Appendixes B and C, we demonstrate this improvement based on hundreds of reconstructions for a range of brightnesses and number of modes. Thus, our method is scalable with the source brightness. 

\section*{Reconstruction of an unknown source }
We use our method to reconstruct mode structure of a mesoscopic parametric down-conversion source, recently reported in \cite{GT}. Bright conjugated beams generated by this source are characterized with TES detectors. The previous work on this source involved using the photon statistics to determine the effective mode number $K$ \cite{Law2004}. However, the nonclassicality characterization based on heralded $g^{(2)}(0)$ quickly saturates with the source brightness. For the dataset analyzed here this criteria provides an inconclusive result. By performing a full JPD analysis, one can affirm the nonclassicality of this bright state and obtain additional information about the types of underlying modes. %and determine the presence of two conjugated thermal modes and a Poissonian noise mode in each arm of the source. 
For a source like the one here, the goal is to engineer a source with a single conjugated mode (with minimal noise), and the information gleaned from the mode reconstruction helps in diagnosing the potential sources of deviation from the ideal case. For applications in which a multimode source is desired, the true number of modes and their brightness (and not just the effective mode number) are required. 
\begin{figure*} [!h!t]
\centering
  \includegraphics[width=0.3\textwidth]{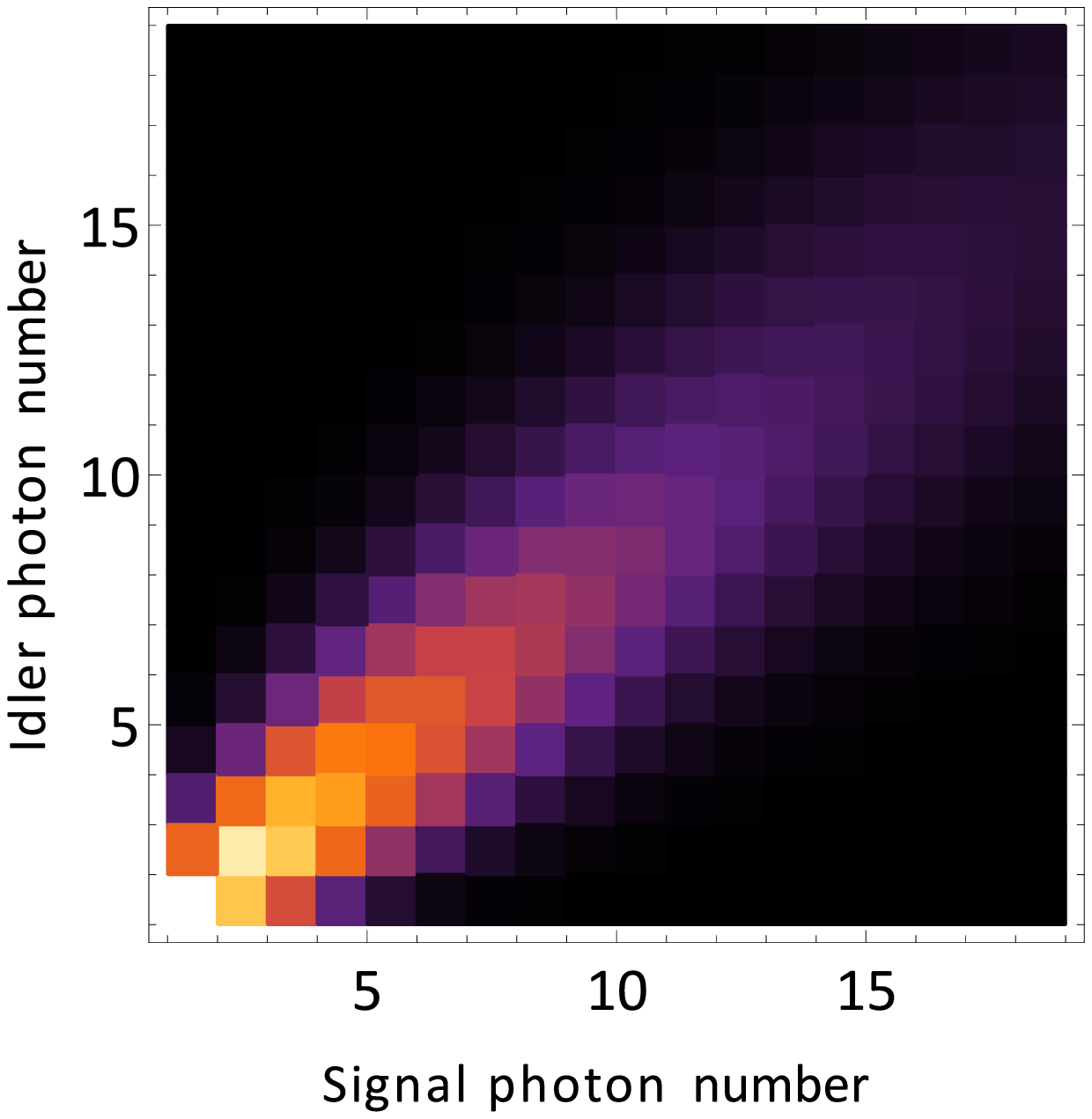}\hspace{0.02\textwidth}
  \includegraphics[width=0.3\textwidth]{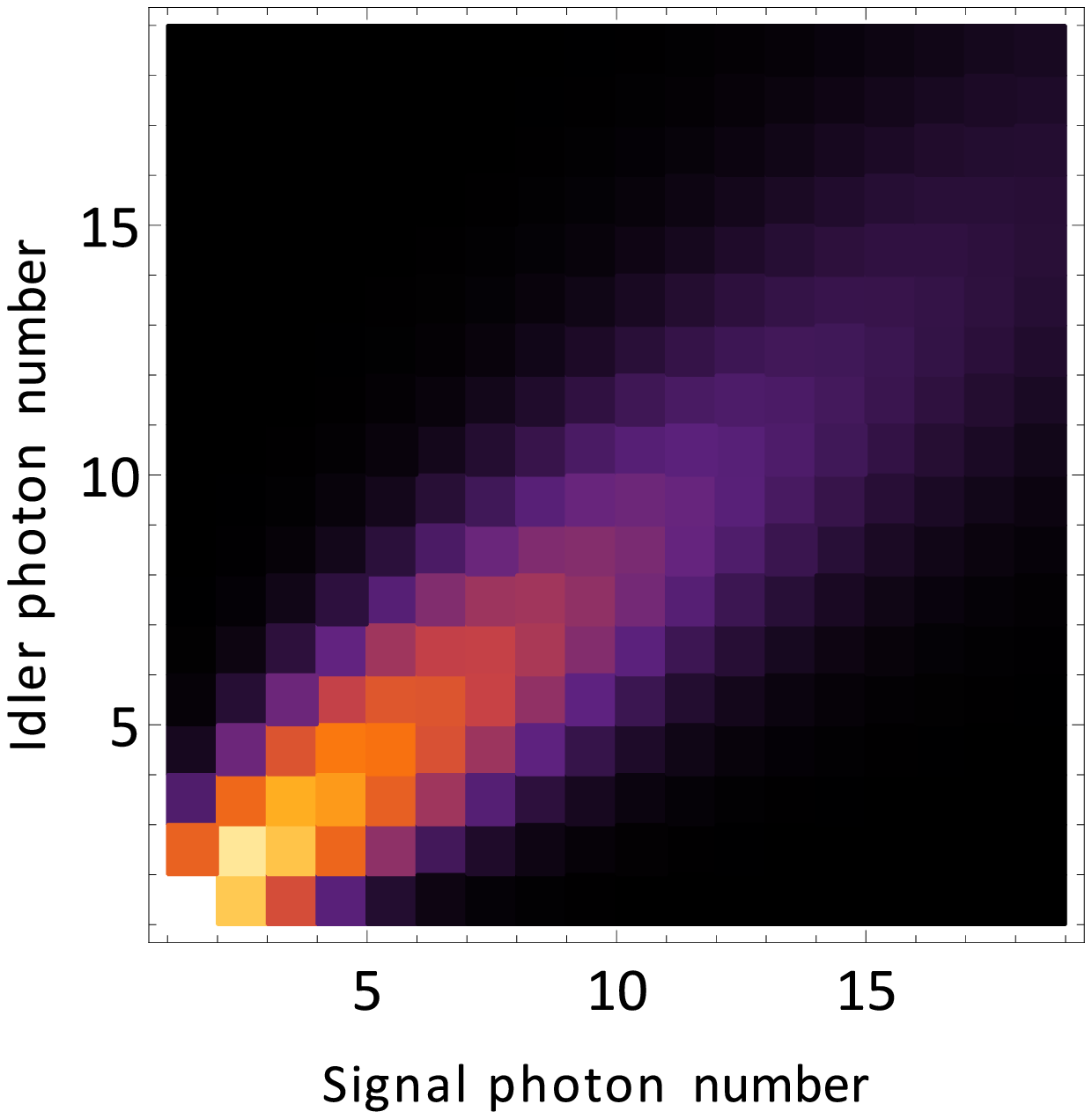}\hspace{0.02\textwidth}
  \includegraphics[width=0.3\textwidth]{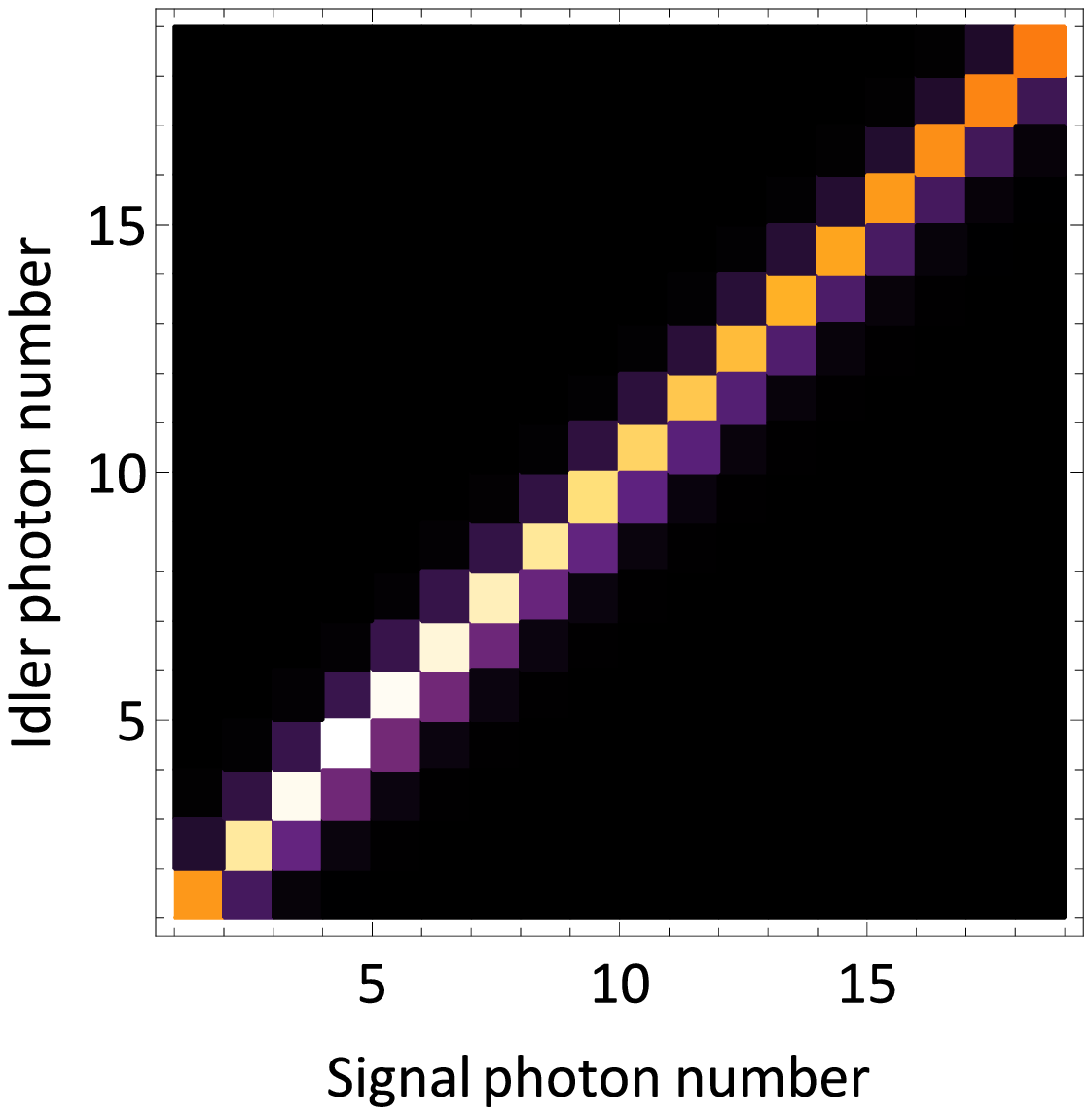}
\caption{\label{fig:two} An experimentally obtained JPD (left), a reconstructed JPD (center) and an initial JPD of a generated light state in the absence of losses. Fit parameters are: $\{$Thermal$_1$, $\mu_1=18.98;\;$Thermal$_2$, $\mu_2=0.93;\;\eta_\mathrm{s}=0.44;\;\eta_\mathrm{i}=0.53;\;$Poisson$_\mathrm{s}$, $\mu_\mathrm{s} = 0.07;\;$Poisson$_\mathrm{i}$, $\mu_\mathrm{i} = 0.21\}$, quality of the fit is characterized by P-value obtained for Pearson's chi-squared test and equals to 0.999999.} 
\end{figure*}

In this reconstruction we use an experimentally found JPD truncated at 40 photons.
In the first step, we calculate RPDs and identify the relevant modes. The number and the type of modes is defined with no prior information. We check for the presence of different mode types using unknown mode analysis  (see Appendixes B), supplemented with uncertainty analysis (see Appendixes C). According to this analysis, we reject all the modes whose population is below $1\%$ of the total mean photon number. Thus, the relevant model contains 2 thermal modes and 1 Poissonian mode per each arm, resulting in two reduced parameter sets $\{$Thermal$_1$, $\mu_1 \eta_\mathrm{s}=8.2;\;$Thermal$_2$, $\mu_2 \eta_\mathrm{s}=0.37;\;$Poisson$_\mathrm{s}$, $\mu_\mathrm{s} \eta_\mathrm{s}=0.1\}$, $\{$Thermal$_1$, $\mu_1 \eta_\mathrm{i}=10.1;\;$Thermal$_2$, $\mu_2 \eta_\mathrm{i}=0.43;\;$Poisson$_\mathrm{s}$, $\mu_\mathrm{i} \eta_\mathrm{i}=0.25\}$. 

In a second step we identify the conjugated modes and losses. Best fitting results are obtained when both thermal modes are considered as conjugated and both Poisson modes as a background, where quality of a fit is assessed using the P-values of Pearson's chi-squared tests.  
For this source, the 2D model (\ref{eqn:JPD}) turns into:
\begin{widetext}
\begin{equation}
p(n_\mathrm{s},n_\mathrm{i})=\underset{n_\mathrm{i}=n^\mathrm{c}_\mathrm{i}+m_\mathrm{i}}{\underset{n_\mathrm{s}=n^\mathrm{c}_\mathrm{s}+m_\mathrm{s}}{\sum}} \left[\underset{n^\mathrm{c}_1,n^\mathrm{c}_2}{\sum}p^\mathrm{Therm}_{\mu^\mathrm{c}_1}(n^\mathrm{c}_1)p^\mathrm{Therm}_{\mu^\mathrm{c}_2}(n^\mathrm{c}_2)L_{n^\mathrm{c}_\mathrm{s},n^\mathrm{c}_1+n^\mathrm{c}_2}(\eta_\mathrm{s})L_{n^\mathrm{c}_\mathrm{i},n^\mathrm{c}_1+n^\mathrm{c}_2}(\eta_\mathrm{i})\right]p^\mathrm{Pois}_{\mu_\mathrm{s}}(m_\mathrm{s})p^\mathrm{Pois}_{\mu_\mathrm{i}}(m_\mathrm{i})
\label{eqn:one}
\end{equation}
\end{widetext}
Using this model, we perform a second stage of mode reconstruction. The experimental JPD, its best fit and a reconstructed JPD that is produced by this source are presented in Fig. \ref{fig:two}. Notice a high quality of this fit, as represented by P-value of 0.999999. We point out that loss figures that are reconstructed through this model with no prior assumptions closely match measured losses in the experiment, and the number and brightness of modes identified here matches the effective number of modes obtained through Schmidt parameter in \cite{GT}. Our reconstruction shows that this source is nearly a single-mode source, and that it has an unprecedented brightness. The reconstructed detection efficiency is in good agreement with the independent experimental calibration. The nonclassicality is confirmed through Hillery criterion ($\sum_{k=1}^{\infty} \sum_{n=0}^{2k}P(n, 2k - n)\leq\sum_{k=0}^{\infty} \sum_{n=0}^{2k+1}P(n, 2k+1 - n)$ for classical states), \cite{Hillery}, on a reconstructed JPD and gives $0.65(4)\nleq 0.34(2)$, i.e. nearly 7 standard deviations \cite{uncertaintynote}. The application of the same criterion on a measured dataset gives $0.4775(5)\leq0.4924(5)$, i.e. the nonclassicality of the original state cannot be directly assessed, in this case due to limited detection efficiency.
\section*{Conclusions}
In conclusion, we have presented a method for reconstruction of a mode structure from joint photon number statistics of a conjugated source and shown that it scales favorably with the source brightness. This method is built upon a model that includes all physically plausible mode statistics applicable to a given physical system, here thermal, Poissonian, and single-photon modes. Beyond the model, this method requires no prior assumptions. In the first step, an experimentally measured JPD for a conjugated source is reduced to two RPDs, each describing the mode structure of one arm of a conjugated field, and a mode structure analysis is performed for each arm separately. In the second step, a mode model of a conjugated field is refined using the results of the first step. A fit to this model determines correlated modes and losses. %A numerical analysis of the first step shows how to identify the number and types of modes present in a field if high-order photon-number distribution data is available. We numerically demonstrate the uncertainty reduction of the second step with the number of experimental measurements performed to estimate JPD. 
We applied our method to experimental JPD data obtained in characterizing an extremely bright parametric down-conversion source, proven its nonclassicality and identified its mode structure, reconstructing all significant modes occupation and type. This is a powerful method, because the only limitation here is that we truncate the mode structure at $\approx1$\% in power - which matches with the expected accuracy of reconstruction due to the experimentally introduced shot noise. Even though a model is needed for a reconstruction, we have shown that our method can build such a model with no prior assumptions. To our knowledge, this is the only method for characterization of mesoscopic nonclassical light. 

\section*{Appendix A: Joint Probability Distribution of a multimode light source}

Consider a lossless case first. The probability to generate $k$ photons in a mode $p_{\mu}(k)$ is governed by that mode's statistics and its mean photon number $\mu$.
Because numbers of photons that occur in each mode are independent, the probability to simultaneously generate $k_j$ photons in each of the mode $j$ is given by a probability multiplication rule: $$P(k_1,...k_j...)=\underset{j}{\prod}p_{{\mu}_j}(k_j).$$ We are interested in finding a probability of generating a total of $M$ photons. The additive rule of probability of mutually exclusive events gives: 
$$P(M)={\underset{\sum{k_j}=M} {\sum}} \underset{j}{\prod}p_{{\mu}_j}(k_j).$$
The conjugated fields ``s'' (signal) and ``i'' (idler) are comprised of correlated ``c'' and uncorrelated ``u'' modes. For the probability to generate $M$ photons in all uncorrelated modes one writes:
$$P_\mathrm{u_{s,i}}(M)={\underset{\sum{k_j}=M} {\sum}} \underset{j}{\prod}p_{{\mu}_j}(k_j).$$
Photons in correlated modes are generated in pairs: the same number in both ``s'' and ``i'' fields: $N_\mathrm{s}=N_\mathrm{i}$. Similarly,
$$P_\mathrm{c}(N_\mathrm{s},N_\mathrm{i})={\underset{\sum{k_j}=N_\mathrm{s}=N_\mathrm{i}} {\sum}} \underset{j}{\prod}p_{{\mu}_j}(k_j).$$
Then, the joint probability distribution is given by applying probability multiplication and additive rules:
\begin{equation}
P(n_\mathrm{s},n_\mathrm{i})=\underset{N_\mathrm{i}+M_\mathrm{i}=n_\mathrm{i}}{\underset{N_\mathrm{s}+M_\mathrm{s}=n_\mathrm{s}}{\sum}} P_\mathrm{c}(N_\mathrm{s}, N_\mathrm{i})P_{\mathrm{u}_\mathrm{s}}(M_\mathrm{s})P_{\mathrm{u}_\mathrm{i}}(M_\mathrm{i}).
\label{one}
\end{equation}
This result can be generalized for the case of losses. A typical procedure to include loss in quantum optics is to act on a state with a beam splitter \cite{MandelBook}. Loss factors $L$ can be written in form of binomial coefficients $L_{n,k}(\eta)=\eta^n(1-\eta)^{k-n}k!/((k-n)!n!)$, where $\eta$ is efficiency (beam splitter transmittance), $k$ is the number of impinging photons and $n$ is the number of photons at the output of this beam splitter. 
\begin{equation}
P_\mathrm{u_{s,i}}(M)=\overset{\infty}{\underset{k=M}{\sum}}{\underset{\Sigma m_j=M}{\underset{\Sigma{k_j}=k} {\sum}}} \underset{j}{\prod}p_{\mu_j}(k_j)L_{m_j,k_j}(\eta_j).
\label{twoprime}
\end{equation}
It can be shown \cite{MandelBook} that losses applied to photon states represented by Poissonian, Binomial and Bose-Einstein statistics do not change their statistics, only affecting the mean photon number in that mode. This allows to introduce adjusted mean photon numbers for uncorrelated modes, significantly simplifying the master equation set:
\begin{equation}
P_\mathrm{u_{s,i}}(M)={\underset{\sum{k_j}=M} {\sum}} \underset{j}{\prod}p_{\tilde{\mu}_j}(k_j),
\label{two}
\end{equation}
where $\tilde{\mu}_j=\mu_j\eta_j$ is an effective mean photon number per mode,  and $\eta_j$ denotes losses in each mode.
For correlated modes, losses in each arm should be independently considered, therefore introducing effective mean photon numbers is no longer possible:
\begin{equation}
P_\mathrm{c}(N_\mathrm{s},N_\mathrm{i})=\overset{\infty}{\underset{(N_\mathrm{s}, N_\mathrm{i})}{\underset{\mathrm{Max}}{\underset{k=}{\sum}}}}\underset{\Sigma n^\mathrm{i}_j=N_\mathrm{i}}{\underset{\Sigma n^\mathrm{s}_j=N_\mathrm{s}}{\underset{\Sigma k_j=k} {\sum}}} \underset{j}{\prod}p_{\mu_j}(k_j)L_{n^\mathrm{s}_j,k_j}(\eta^\mathrm{s}_j)L_{n^\mathrm{i}_j,k_j}(\eta^\mathrm{i}_j).
\label{three}
\end{equation}
Equation \ref{one} is not affected by the losses. Generally speaking, Eqs. (\ref{one}), (\ref{twoprime}), (\ref{three}) describe a mode mixing with all types of modes, i.e with any underlying statistical distribution. Because all mesoscopic optical states demonstrated to date are comprised of the modes with Poissonian, Binomial and Bose-Einstein statistics only, we take advantage of the simplification outlined above. Thus, (\ref{one}), (\ref{two}), and (\ref{three}) give Eq. (1) in the main manuscript.

\section*{Appendix B: Unknown source types}

It was established \cite{PolyGold} that a 1D reconstruction through photon-number statistics works well when number of modes and their types are known beforehand. Furthermore, adding unoccupied modes to the reconstruction only requires expanding experimental data sets to achieve the same accuracy, but otherwise does not negatively affect the reconstruction. However, not including all modes present in a reconstruction leads to significant errors in the entire set of recovered parameters. In the most general case, it is useful to establish a method to identify an {\it a priori} unknown mode structure based on a series of reconstructions, a situation typical for the experimental data. 

Here we present the details of how to reconstruct such mode structure. To aid this task we use recovered sets of fitting parameters $\mathcal{S}_\mathrm{recovered}$ together with their corresponding absolute fitting errors, $\sqrt{\sum(x_j-f_j)^2}$, where $x_j$ represents the simulated probability distribution and $f_j$ represents the recovered probability distribution.  

In most cases, it is important to determine if a Poisson mode is present, or a distribution can be well-described by a finite number of thermal modes. To show this, we generate a Poisson probability distribution and fit it multiple times, with an increasing number of thermal modes. The results are presented in Table ~\ref{tab:one}. We see that all the fits yield equal population of all available thermal modes. In hindsight, it is not surprising, because a maximally uncorrelated state with N thermal modes occurs when all the modes are equally populated \cite{MandelBook}. Therefore, if reconstructions yield mode populations that depend on the number of thermal modes allowed, the number of thermal modes in the reconstruction should be increased (and a Poissonian mode allowed) until the reconstruction no longer depends on the number of modes allowed. The size of the experimental data set, including both the maximum photon number detected and the amount of data, will ultimately limit the number of modes that can be included in an accurate reconstruction. 
 
\begin{table}
\centering
\caption{\bf Poisson mode with $\langle n\rangle=5$ fitting with multiple thermal modes}
\label{tab:one}
\begin{tabular}{|c|c|c|c|}
\hline
Number of modes & Photons per mode & $\langle n\rangle$ & Error $\sqrt{\sum(x_j-f_j)^2}$ \\
\hline
1 & 1.82 & 1.82 & 0.47\\
2 & 1.3 & 2.6 & 0.33\\
3 & 1.03 & 3.03 & 0.26\\
4 & 0.85 & 3.4 & 0.22\\
10 & 0.42 & 4.2 & 0.11\\
20 & 0.23 & 4.6 & 0.06\\
40 & 0.12 & 4.8 & 0.03\\
80 & 0.061 & 4.88 & 0.017\\
\hline
\end{tabular}
\end{table}

\section*{Appendix C: The effect of uncertainty in a JPD measurement}
\begin{figure}
\centering
   \includegraphics[width=0.95\columnwidth]{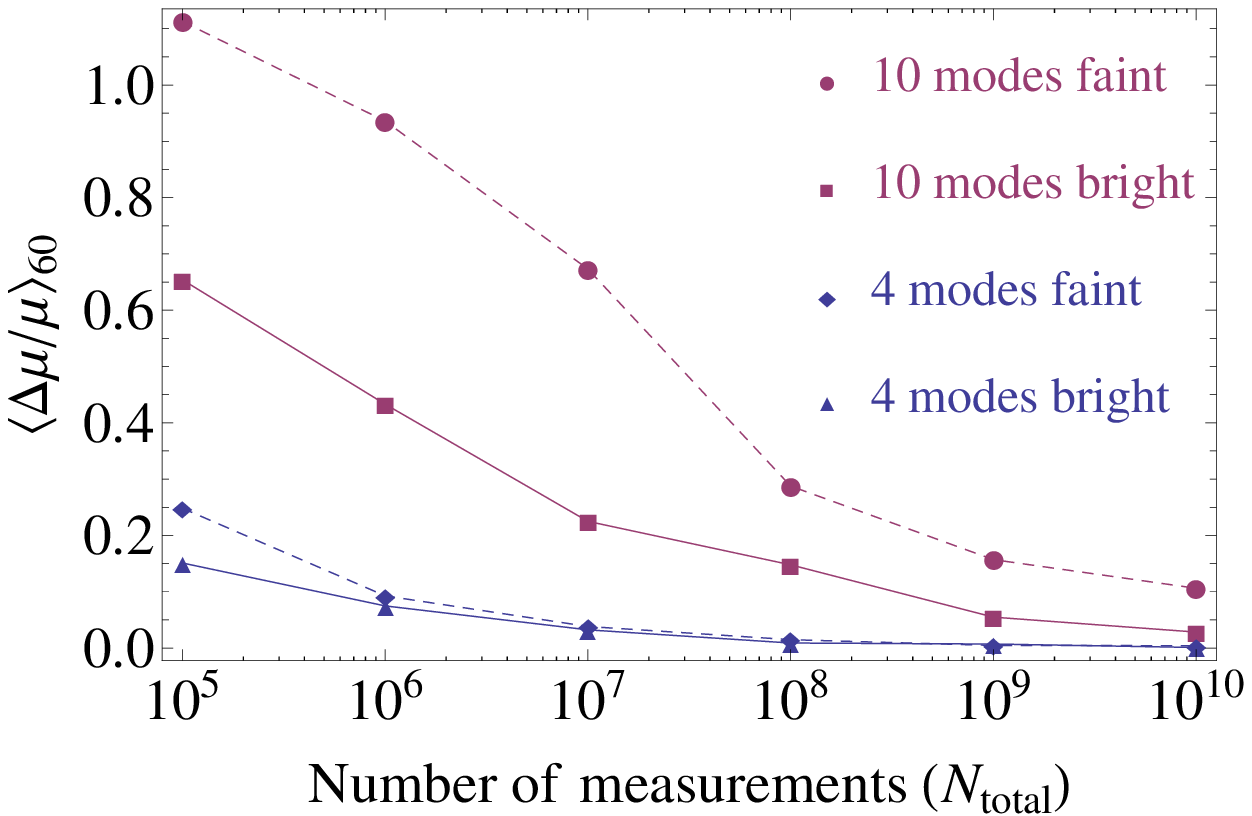}
  \caption{\label{fig:noise} Relative fitting error vs. number of experimental trials. Bright 4-mode field reconstruction with $\{$Thermal$_1$, $\mu_1=19;\;$Thermal$_2$, $\mu_2=1;\;\eta_\mathrm{s}=0.43;\;\eta_\mathrm{i}=0.52;\;$Poisson$_\mathrm{s}$, $\mu_\mathrm{s} = 0.1;\;$Poisson$_\mathrm{i}$, $\mu_\mathrm{i} = 0.2\}$  (blue triangles); Bright 10-mode field reconstruction with $\{$Thermal$^\mathrm{c}_1$, $\mu^\mathrm{c}_1=4;\;$Thermal$^\mathrm{c}_2$, $\mu^\mathrm{c}_2=3;\;$Single Photon$^\mathrm{c}_3$, $\mu^\mathrm{c}_3=0.6;\;$Poisson$^\mathrm{c}_4$, $\mu^\mathrm{c}_4 = 1;\;\eta_\mathrm{s}=0.5;\;\eta_\mathrm{i}=0.6;\;$Thermal$^\mathrm{s}_1$, $\mu^\mathrm{s}_1 = 1.5;\;$Thermal$^\mathrm{s}_1$, $\mu^\mathrm{s}_2 = 1;\;$Single Photon$^\mathrm{s}_3$, $\mu^\mathrm{s}_3 = 0.8;\;$Thermal$^\mathrm{s}_1$, $\mu^\mathrm{i}_1 = 2;\;$Thermal$^\mathrm{i}_1$, $\mu^\mathrm{i}_2 = 0.4;\;$Single Photon$^\mathrm{i}_3$, $\mu^\mathrm{i}_3 = 0.7\}$ (purple squares). Faint light sources have 10-fold lower photon numbers of corresponding bright sources (blue diamonds: 4-mode field; purple circles: 10-mode field). Lines are guides to an eye.}
\end{figure}
Shot noise must always be considered in analyzing the accuracy of a method for characterization based on optical detection. Additionally, truncating detection at some photon number limits the accuracy of the extracted photon number probabilities. We see that this method, with detection of photon numbers larger than the number of modes present, is robust against shot noise over a range of total mean photon number. Typically, the uncertainty for $p(n_\mathrm{s}, n_\mathrm{i})$ where $n_\mathrm{s}$, $n_\mathrm{i}$ photons are detected in the idler, signal arms respectively can be estimated from the shot-noise limit under the assumption that trials are independent. If $N(n_\mathrm{s}, n_\mathrm{i})$ is the number of observed events, then the uncertainty is given by $\sqrt{N(n_\mathrm{s}, n_\mathrm{i})}$. Then, for each element of JPD $p(n_\mathrm{s}, n_\mathrm{i})$ the absolute uncertainty is $\sqrt{p(n_\mathrm{s}, n_\mathrm{i})/N_\mathrm{tot}}$, where $N_\mathrm{tot}=\underset{n_\mathrm{s},n_\mathrm{i}}{\sum}N(n_\mathrm{s}, n_\mathrm{i})$ is the total number of trials. Using numerical simulations, we investigate the effect of experimental uncertainty on reconstruction quality. We simulate experimentally obtained JPDs by computing an exact JPD and adding random shot noise corresponding to a different number of trials $N_\mathrm{tot}$. Then we calculated a relative reconstruction error (averaged over 60 independently seeded, random shot noise sets) as a deviation of reconstructed mode brightness from those known exactly, $\langle\Delta\mu/\mu\rangle_{60}=\langle\sum\left|\mu_j^\mathrm{rec}-\mu_j^\mathrm{sim}\right|\rangle_{60}/\sum\mu_j^\mathrm{sim}$. Typical dependencies of the relative reconstruction error on number of trials for a 4-mode and a 10-mode fields of different brightness are shown in Fig. \ref{fig:noise}. Note that field parameters of the 4-mode bright field in this figure are similar to what we expect in our experimental data, see below; a 10-mode bright field reconstruction is simulated for a field comprised of the same modes and with the same losses as that considered in Fig. 1 of the manuscript. Next, Fig. \ref{fig:noise} demonstrates that the accuracy of our method improves with the field brightness, i.e. that our method is scalable to macroscopic states of light. Both faint and bright reconstructions of 4- and 10- mode fields have comparable mean photon number and losses pairwise, thereby the demonstrated difference in accuracy is primarily due to the number of fit parameters. As expected, uncertainty decrease monotonically with the number of trials. This uncertainty analysis plays an important role in establishing a number of trials required to reach a desired accuracy of reconstruction. This analysis also sets a limit on the number of modes in a reconstruction model, because unoccupied modes will be populated to within the reconstruction accuracy and should be excluded from a model. 

\bibliography{biblio}

\begin{thebibliography}{30}
\bibitem[{unc()}]{uncertaintynote}
\bibinfo{note}{Uncertainty is estimated from the uncertainty on a reconstructed
  mean photon number}.
\end{thebibliography}

\end{document}